\documentclass[a4paper,twocolumn,superscriptaddress]{revtex4-1}

\usepackage{hyperref}
\usepackage{multirow}
\usepackage{xcolor}
\hypersetup{
	colorlinks,
	linkcolor={red!90!black},
	citecolor={black!10!blue},
	urlcolor={blue!80!black}
}

\usepackage{bm,amsmath}
\usepackage{amssymb}
\usepackage{graphicx}
\usepackage{array}
\newcommand{\f}{\frac}
\newcommand{\ud}{\uparrow\to\downarrow}
\newcommand{\du}{\downarrow\to\uparrow}
\newcommand{\al}{\langle \ell\rangle}

\newcommand{\mZ}{\mathcal{Z}}

\begin{document}

\title{Stretched exponential to power-law: crossover of relaxation in a kinetically constrained model }

\author{Sukanta Mukherjee}
\affiliation{Tata Institute of Fundamental Research, Hyderabad - 500046, India}
\affiliation{Cluster of Excellence Physics of Life, Technische Universität Dresden, Arnoldstraße 18, 01307 Dresden, Germany}

\author{Puneet Pareek}
\affiliation{Tata Institute of Fundamental Research, Hyderabad - 500046, India}

\author{Mustansir Barma}
\email{barma@tifrh.res.in}
\affiliation{Tata Institute of Fundamental Research, Hyderabad - 500046, India}

\author{Saroj Kumar Nandi}
\email{saroj@tifrh.res.in}
\affiliation{Tata Institute of Fundamental Research, Hyderabad - 500046, India}

\begin{abstract}
The autocorrelation function in many complex systems shows a crossover in the form of its decay: from stretched exponential relaxation (SER) at short times to power law at long times. Studies of the mechanisms leading to such multiple relaxation patterns are rare. Additionally, the inherent complexity of these systems makes it hard to understand the underlying mechanism leading to the crossover. Here we develop a simple one-dimensional spin model, which we call a Domain Wall (DW) to Doublon model, that shows such a crossover as the nature of the excitations governing the relaxation dynamics changes with temperature and time. The relevant excitations are DWs and bound pairs of DWs, which we term `doublons'. The diffusive motion of the DWs govern the relaxation at short times, whereas the diffusive motion of the doublons yields the long time decay. This change of excitations and their relaxation leads to a crossover from SER to power law in the decay pattern of the autocorrelation function. We augment our numerical results with simple physical arguments and analytic derivations. 
\end{abstract}

\maketitle

%Relaxation phenomena, characterized via the decay of time autocorrelation functions, in complex-condensed matter systems can specify how microscopic dynamics are related to the macroscopic observations through the decay laws. 

Relaxation dynamics is fundamental in various branches of science, ranging from the kinetics of reaction rates  \cite{weiss1986} to diffusion in a complex environment \cite{hanggi1990} to escape problems \cite{kramers1940}. For simple systems, in the absence of disorder or trapping, barrier crossing between valleys leads to exponential relaxation, known as the Maxwell-Debye relaxation \cite{maxwell1867}. However, the presence of disorder or spatial heterogeneity leads to a more complex scenario, often resulting in {\it stretched exponential relaxation (SER)}, as observed in different systems. Examples include dynamics in glasses \cite{giulioreview,phillips1996}, dielectric relaxation in many organic compounds \cite{glarum1960,skinner1983}, systems with arrested steady states \cite{das1999,vaibhav2020}, etc. It is now well-known that SER can arise from many different mechanisms (see Ref. \cite{vaibhav2020} for a brief overview). For example, Glarum \cite{glarum1960} and Bordewijk \cite{bordewijk1975} studied the diffusion dynamics of vacancies in liquid structure to obtain SER. Skinner \cite{skinner1983}, Spohn \cite{spohn1989}, Godr{\`{e}}che \cite{godreche2015}, and others \cite{ritort2003} have extended such frameworks to diffusive dynamics of domain walls in spin systems. A model with competing interactions shows SER with the unusual feature that the relaxation time diverges as system size increases \cite{vaibhav2020}. On the other hand, the relaxation in several complex systems, such as spin glasses below the transition temperature \cite{maclaughlin1983,keren1996}, systems close to a critical point \cite{chaikinlubensky,stanleybook}, stress relaxation in block copolymers \cite{rubenstein1993}, are described by even slower decays, characterized by {\em power laws}.

Most systems generally follow a single relaxation form that describes the decay of the correlation functions. However, there are several systems that show crossovers from one type of decay of the correlation function to another as time progresses. Examples include the dynamics in non-entangled polymeric melts \cite{ganazzoli2002}, survival kinetics in random walks with fractally correlated traps \cite{plyukhin2016}, the ensemble-averaged relaxation process in a model with arrested states \cite{vaibhav2020}, the angular-velocity autocorrelation function of an active probe in a glassy medium \cite{clara2020}, the self-overlap function in particulate \cite{pareek2023} and confluent biological systems \cite{souvik2020,pandey2023}, protein conformational dynamics \cite{yang2002,yang2003}, etc. The autocorrelation function in these systems shows a crossover from an SER form to a power law at later times. Although many works have focused on understanding the mechanisms leading to either SER or power law, detailed studies of the dynamics where both these forms are present and the mechanism leading to the crossover are relatively rare.

Metzler, Barkai, and Klafter \cite{metzler1999} introduced a fractional Fokker-Planck equation describing particle dynamics under external and thermal driving and generalized the standard Markoffian model of diffusive dynamics to fractional dynamics for disordered systems \cite{metzler2000,barkai2001,metzler2002,burov2008a,lizana2010}. They demonstrated that relaxations in such systems follow a Mittag-Leffler function \cite{mainardi2020}, which shows an initial SER followed by a power law at long times. An interesting result in this context is that the stretching exponent, $\beta$, of the SER: $C(t)\sim \exp[-(t/\tau)^\beta]$, where $t$ is time and $\tau$ is a relaxation time, and the exponent, $\alpha$, which characterizes the power law: $C(t)\sim (t/\tau)^{-\alpha}$, are the same, that is $\alpha=\beta$.
However, there are systems where $\alpha$ and $\beta$ can differ, for example, the cellular Potts model in $2D$ \cite{souvik2020} or systems with more complexity, which was also hypothesized in Ref. \cite{metzler2002}. Whether and how $\alpha$ should be related to $\beta$, in general, remains unclear. Since the well-known systems discussed above showing two distinct forms for the decay of the autocorrelation function at different times are complex, it is advantageous to have simple model systems showing similar dynamical characteristics, in order to draw meaningful insights.

\begin{figure*}
	\includegraphics[width=12.6cm]{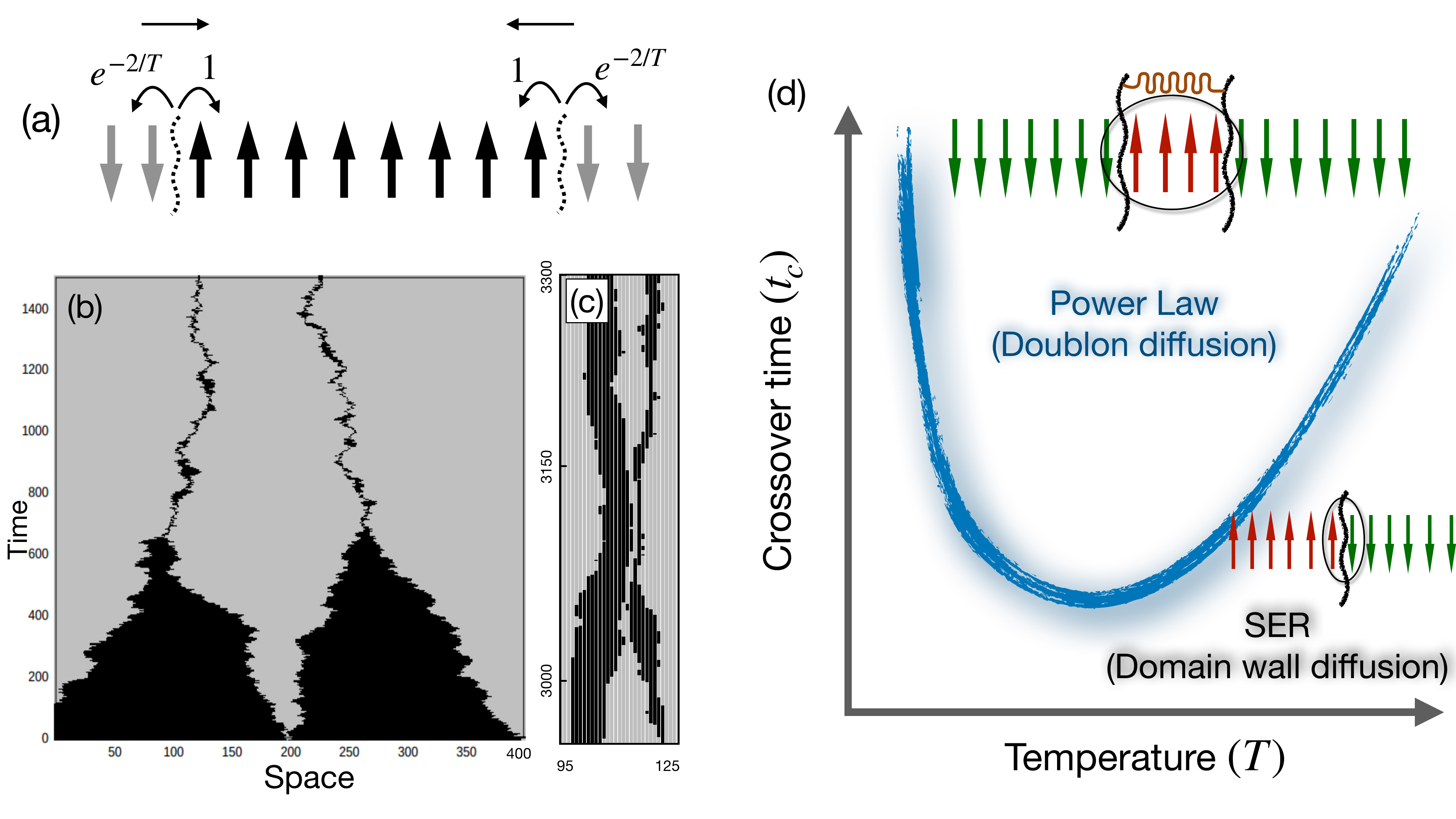}
	\caption{Basic phenomenology of the dynamics. (a) Consider a domain of $\uparrow$ spins surrounded by $\downarrow$ spins. Due to the energy function, Eq. (\ref{onsiteenergy}), the probability of flipping a $\uparrow$ spin is $1$, whereas that for a $\downarrow$ spin is $e^{-2/T}$. The constraint of the dynamics allows spin-flip only at the boundary of a domain. Thus, the movement of DWs towards the $\uparrow$ spins is favored, as indicated by the arrows. This leads to an effective attractive interaction between DWs separated by $\uparrow$ spins. Such a pair of DWs form a bound state that we term a doublon. (b) Evolution of a system with four DWs, created by two $\downarrow$ spins (marked gray) at equal distances in a system of $\uparrow$ spins (marked black). The system size is $400$. The dynamics leads to the formation of two doublons at long times. (c) Hard-core repulsion between two doublons in our simulation. (d) Schematic representation of $t_c$ as a function of $T$. The crossover time, $t_c$, diverges both for $T\to0$ and $T\to\infty$, for different reasons. The crossover time separates the power-law decay from a faster decay (the SER at higher $T$) of the autocorrelation function. We emphasize that the crossover time is not sharply defined. The relevant excitations in the two regimes are schematically shown: domain wall in the SER regime and doublon (bound states of two domain walls) in the power law regime. }
	\label{schematic_dyn}
\end{figure*}

In this work, we study a simple one-dimensional spin model, which we term a Domain Wall to Doublon (DWD) model. In a large range of temperature, it shows a crossover of dynamics from SER at short times to a power-law form at long times. Motivated by two-level systems of resonantly driven Rydberg atoms in optical lattices \cite{browaeys2020}, Causer {\it et al.} \cite{Causer2020} first introduced and studied this specific model under the name of the XOR-FA model; however, the focus of that study was different from ours. We find that the relaxation dynamics in the DWD model is governed by excitations whose character changes in time: at short times, the relevant excitations are individual domain walls (DWs), while at long times, they are bound pairs of DWs that we call ``doublons". The number of DWs in our model is conserved, and there is a hard-core repulsion between them. The short-time dynamics involves individual DW motion, leading ultimately to the formation of doublons, as adjacent pairs of DWs move towards each other. On the other hand, the long-time dynamics comprises the diffusive motion of the doublons. A crossover in the form of the dominant excitation leads to the crossover of the decay pattern of the autocorrelation function; at high temperature, this is a crossover from stretched exponential to power law decay.

 The organization of the rest of the paper is as follows: we introduce the model in Sec. \ref{modeldescription}.  For the results, we first present the steady state properties in Sec. \ref{results1} and then the details of the dynamics in Sec. \ref{results2}. We conclude the paper in Sec. \ref{disc} with a brief discussion of our results.

\section{Description of the model}
\label{modeldescription}
The DWD model defined below is a variant of a specific class known as kinetically constrained models (KCMs). Such models have given crucial insights into the dynamics of glassy systems \cite{ritort2003,garrahan2002,katira2019}. They are also relevant to operator spreading and dynamics in quantum many body systems \cite{nahum2017,banuls2019}, dynamics of Rydberg atoms \cite{lesanovsky2013}, gravity-driven granular flow and clogging \cite{bolshak2019}, etc.

The DWD model is defined on a one-dimensional lattice with an Ising spin $S_i=\pm 1$ on each site $i$. The system is governed by an on-site energy function,
\begin{equation} \label{onsiteenergy}
\mathcal{H}=\sum_{i=1}^N S_i,
\end{equation}
where $N$ is the total number of spins. Unless otherwise mentioned, we use periodic boundary conditions. Equation (\ref{onsiteenergy}) describes non-interacting spins in a magnetic field of unit strength. The system evolves through single spin-flip dynamics with a kinetic constraint: a spin can flip only if its two nearest neighbors have opposite signs. This latter constraint ensures that the number of DWs remains constant; an allowed spin-flip results in the DW moving one lattice spacing. Thus the space of all configurations is partitioned into `microcanonical' sectors, each labelled by $N_d$, or equivalently by $\cal E $ $= \sum_i S_i S_{i+1}$ $=(N - 2N_d)$.  

Our Monte Carlo simulation proceeds as follows: at each step, we choose a spin at random; if the two nearest neighbors of this spin are opposite, we flip the spin with probability min$(1,\exp[2S_i/T])$. $N$ such attempts define the unit of time. In other words, the transition rate $W(\uparrow\to\downarrow)$ is unity, while $W(\downarrow\to\uparrow)$ is $c=exp(-2/T)$, ensuring that the condition of detailed balance
\begin{equation}
	\f{W(\du)}{W(\ud)}=c=e^{-2/T}
\end{equation}
holds within each sector. We have set the Boltzmann constant, $k_B$, to unity.

Causer {\it et al.} motivated the constraint on number of DWs as a representation of the dynamics in an ensemble of Rydberg atoms \cite{browaeys2020} in their ``facilitated" states: ``{\it When driven out of resonance, specifically when blue-detuned, conditions can be such that an atom may change state only if a single neighbor is in the excited state, but not both}" \cite{Causer2020}. This physical system would be an ideal candidate to experimentally test the dynamics explored in the current work.

\begin{table*}
	\caption{Comparison of the properties of the three distinct but related models.}
	\begin{tabular}{ | m{2.8cm}| p{3.7cm}|p{2.7cm}|p{5cm}|  }
		\hline
		Models & Domain Wall Conserved & Energy Conserved & Properties \\
		\hline
		Energy Conserving Ising Spins   & Yes    &Yes&   SER, $\beta = 1/2$ \cite{spohn1989}\\
		\hline
		East Model &   No  & No   & SER, $\beta = 1/2$ \cite{sollich2003,ritort2003}\\
		\hline
		DWD Model &Yes & No&  Crossover, SER ($\beta = 1/2$) to power law (Exponent = $1/2$) [This work]\\
		\hline
	\end{tabular}
	\label{table}
\end{table*}

It is instructive to compare our model with a couple of other kinetically constrained models which have been studied extensively. At $T=\infty$, the dynamics of our model becomes identical to the energy-conserving dynamics in an Ising spin model with nearest-neighbor interactions. Such dynamics arises, for example, in equilibrium under a sudden quench from a finite $T$ to $T = 0$ \cite{skinner1983,spohn1989}. The autocorrelation function then follows SER with $\beta=1/2$ in $1D$.

If we were to replace the kinetic constraint with one where spin-flips are allowed only when one of the nearest neighbors is up, the model would become equivalent to another kinetically constrained model \cite{crisanti2000,ritort2003}: Depending on whether the spin flip is allowed if the left (or right) neighbor is up, the model is called the East (or West model), respectively. The autocorrelation function follows SER with $\beta=1/2$ in this case too. 

However, there is a difference between these two classes of models: In the first class, the total energy and number of DWs are both conserved, whereas in the second class both can vary. Both result in SER with $\beta=1/2$, but in our model, we find that the autocorrelation function shows a crossover from SER to a power law, the DW number is conserved, whereas the total energy can change (see Table \ref{table}).

{\it Essential characteristics of the model:}
\label{phenomenology}
Before getting into the detailed results, it is instructive to first look into some qualitative aspects of the model. Due to the kinetic constraint, spin flips are allowed only at the interfaces of two domains, i.e., allowed moves involve movements of the DWs. At $T=\infty$, each DW performs a random walk but with a hard-core constraint between DWs. When $T<\infty$, there is a higher flipping rate for a $\ud$ than $\du$ owing to the applied field.

Now, consider a domain of $\uparrow$ spins surrounded by down spins as schematically shown in Fig. \ref{schematic_dyn}(a). For any finite value of $T$, the transition $\ud$ is more likely than $\du$. Therefore, at both ends of a domain, there is a tendency to shrink the $\uparrow$ spin domain, which is tantamount to an effective attractive interaction between the two DWs surrounding each domain of $\uparrow$ spins; the strength of this interaction is largest at $T=0$ and vanishes as $T \to \infty$.

As a result of this effective attractive interaction at $T < \infty$, DWs at the extremities of a domain of $\uparrow$ spins move towards each other and form a bound state that we define as a ``doublon". Periodic boundary conditions imply an even number, $N_d$, of DWs. Conservation of DW number implies the conservation of the number of doublons, given as $N_d/2$. The minimum size of a doublon is one lattice spacing ($\downarrow\boldsymbol{\uparrow}\downarrow$). There is a hard-core repulsion between different doublons; this comes from the hard-core repulsion among the DWs and the minimum separation between two doublon is two lattice spacings ($\downarrow\boldsymbol{\uparrow}\downarrow\boldsymbol{\uparrow}\downarrow$). At finite $T$ the doublon length fluctuates around an average, $\al$. Interestingly, the diffusion of the doublons dominates the long-time dynamics. 

Figure \ref{schematic_dyn}(b) shows the evolution of a system with four DWs, created by two $\downarrow$ spins at equal distances in a system of $\uparrow$ spins. The DWs between $\uparrow$ spins move toward each other and form doublons. Figure \ref{schematic_dyn}(c) shows the evolution of two such doublons in our simulation. They move away from each other after coming two lattice spacings apart. Figure \ref{schematic_dyn}(d) shows schematically the $T$-dependence of crossover time, $t_c$, beyond which the doublons form and diffuse, thereby determining the long-time dynamics.

\section{Results: static properties}
\label{results1}

\subsection{Analytic theory for domain distribution}
We first provide analytic results for the static properties in the steady state. For a large enough system of size $N$, we may use a grand canonical description, within which the domain wall number is controlled by a conjugate chemical potential. The grand canonical partition function is
\begin{align}
\mZ(\mu,T)&=\sum_{\mathbb{C}} \exp\left[\frac{\mu N_d}{T}\right]\exp\left[{-\frac{1}{T} \sum_{i=1}^N S_i}\right] ,
\end{align}
where $\mu$ is the chemical potential that controls the DW density, and $\mathbb{C}$ represents all possible configurations. Since the number of DWs is given by $N_d=\sum_{i=1}^N (1-S_iS_{i+1})/2$, we obtain
\begin{align}\label{partitionfn1}
\mZ(\mu,T)=e^{\f{\mu N}{2T}}\sum_{\mathbb{C}}e^{ -\frac{\mu}{2T}\sum_{i=1}^NS_i S_{i+1}-\frac{1}{T}\sum_{i=1}^NS_i}.
\end{align}
Observing that Eq. (\ref{partitionfn1}) is just the partition function of the 1D nearest neighbor Ising model, we may use the standard transfer matrix formalism \cite{pathriabook} to obtain 
\begin{equation}
\mZ(\mu,T)=\Lambda_{+}^N+\Lambda_{-}^N
\end{equation}
where
\begin{align}\label{partitionfunc}
\Lambda_{\pm}=\left[\cosh(1/T)\pm\sqrt{z^2+\sinh^2(1/T)}\right],
\end{align}
with $z=\exp[\mu/T]$. For large $N$, the first term, $\Lambda_+^N$, dominates.

Let us define an $\ell$-cluster as a stretch of $\ell$ consecutive up spins, with down spins at the two ends. The probability $P(\ell)$ of the occurance of an $\ell$-cluster is
\begin{equation}
P(\ell)= \langle \downarrow\uparrow\uparrow \ldots \uparrow\downarrow\rangle= \langle \bar{n}_i {n}_{i+1} {n}_{i+2}\ldots {n}_{i+\ell}\bar{n}_{i+\ell+1}\rangle,
\end{equation}
where $n_i$ and $\bar{n}_i$ are occupancy variables for $\uparrow$ and $\downarrow$ spins respectively, and are given by
\begin{equation}
n_i=\frac{1+S_i}{2}; \,\,\, \bar{n}_i=\frac{1-S_i}{2}.
\end{equation}

After straightforward algebra, we obtain
\begin{align}\label{theorydist}
P(\ell)=C \f{\exp(-\f{\ell}{T})}{\left[\cosh(\f{1}{T})+\sqrt{z^2+\sinh^2(\f{1}{T})}\right]^{\ell+2}},
\end{align}
where $C$ is given as
\begin{align}
C=\frac{z^4e^{-\f{1}{T}}\left[e^\f{1}{T}+\sinh(\f{1}{T})+\sqrt{z^2+\sinh^2(\f{1}{T})}\right]}{z^2+\left\{ \sinh(\f{1}{T})+\sqrt{z^2+\sinh^2(\f{1}{T})} \right\}^2} \nonumber.
\end{align}

Further, the DW density, $\rho_{DW}$, is given by
\begin{equation}\label{dwdensity}
\rho_{DW}=\frac{\langle N_d\rangle}{N}=\frac{z}{N}\frac{\partial \mathrm{ln}\mZ(\mu,T)}{\partial z}\approx \f{z}{N}\f{\partial \Lambda_{+}}{\partial z}.
\end{equation}
Using Eq. (\ref{partitionfunc}), we obtain
\begin{equation}\label{chempot}
\rho_{DW}=\frac{z^2}{z^2+\sinh^2(\f{1}{T})+ \cosh(\f{1}{T})\sqrt{z^2+\sinh^2(\f{1}{T})}}.
%{e^{\f{2\mu}{k_BT}}+\sinh^2(\f{1}{k_BT})+\cosh(\f{1}{k_BT})\sqrt{e^{\f{2\mu}{k_BT}}+\sinh^2(\f{1}{k_BT})}}
\end{equation}
For a given $\rho_{DW}$, we obtain $z$ using Eq. (\ref{chempot}) at a particular $T$. We then calculate $P(\ell)$ at different $T$ using Eq. (\ref{theorydist}).

\begin{figure}
	\includegraphics[width=8.6cm]{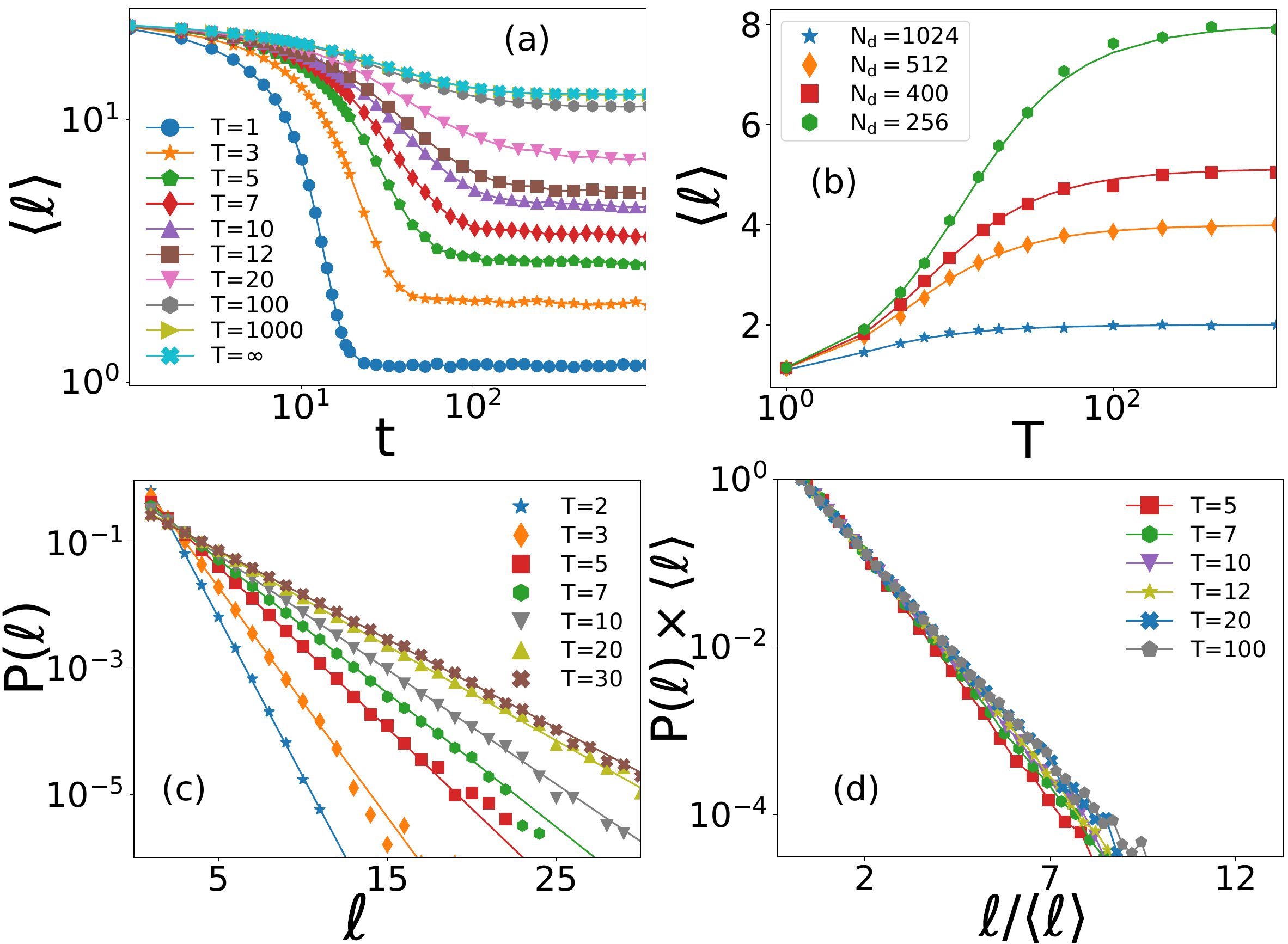}
    \caption{(a)	Evolution of average doublon length, $\al$, with 160 equally spaced $\downarrow$ spins and the rest being $\uparrow$ spins with system size $N=4000$. $\al$ attains a $T$-dependent equilibrium value. Each point is an average of $10^4$ systems. (b) Equilibrium behaviour of $\al$ as a function of $T$, with $N=2048$ and different numbers, $N_d$, of DWs, each point is averaged over $10^4$ systems. $\al$ saturates to the average domain length, $N/N_d$, at high $T$, and is independent of system size at low $T$. The lines are from theory, obtained via Eq. (\ref{theorydist}), and the symbols are simulation data. (c) The distribution of domain lengths, $P(\ell)$, follows an exponential trend. Lines are from theory (Eq. \ref{theorydist}), and symbols are simulation data. We have used $N=1024$ and $\rho_{DW}=0.25$ in the simulations. (d) The plot of the domain length distribution function as a function of $\ell/\al$ follows a master curve. }
	\label{mfpt_domain}
\end{figure}

\subsection{Simulation results for steady-state domain distribution}

We now present the simulation results for the steady-state doublon length, $\ell$, defined as the number of $\uparrow$ spins between two consecutive DWs. We first simulate a system of fixed size with periodic boundary conditions and a certain number of $\downarrow$ spins (each of which creates two DWs) in a system of $\uparrow$ spins. We start with an initial condition where the $\downarrow$ spins are equally spaced. As the system evolves, $\ell$ reaches its equilibrium distribution. The evolution of the average doublon length, $\al$, as a function of time, is shown in Fig. \ref{mfpt_domain}(a) for different values of $T$. At $T=\infty$, there is no difference between the domains of $\uparrow$ or $\downarrow$ spins, and $\al$ is simply the average domain length given by $N/N_d$ where $N$ is the size of the system and $N_d$ is the conserved number of DWs. As shown in Fig. \ref{mfpt_domain}(b), $\al$ monotonically decreases with decreasing $T$ and becomes unity (one $\uparrow$ spin) at $T=0$. The analytic result for $\al$ using Eq. (\ref{theorydist}) is shown by continuous lines in Fig. \ref{mfpt_domain}(b).

Fig. \ref{mfpt_domain}(c) shows that the distribution $P(\ell)$ at different $T$ agrees well with Eq. (\ref{theorydist}). The decay of $P(\ell)$ is slower at higher $T$; as $\al$ grows monotonically as $T$ increases. To confirm that $P(\ell)$ follows a nearly exponential form, $P(\ell)\sim \exp[-\ell/\al]$, we plot $\al P(\ell)$ as a function of $\ell/\al$ in Fig. \ref{mfpt_domain}(d) and find all data follow a master curve as expected.

\begin{figure}
\includegraphics[width=8.6cm]{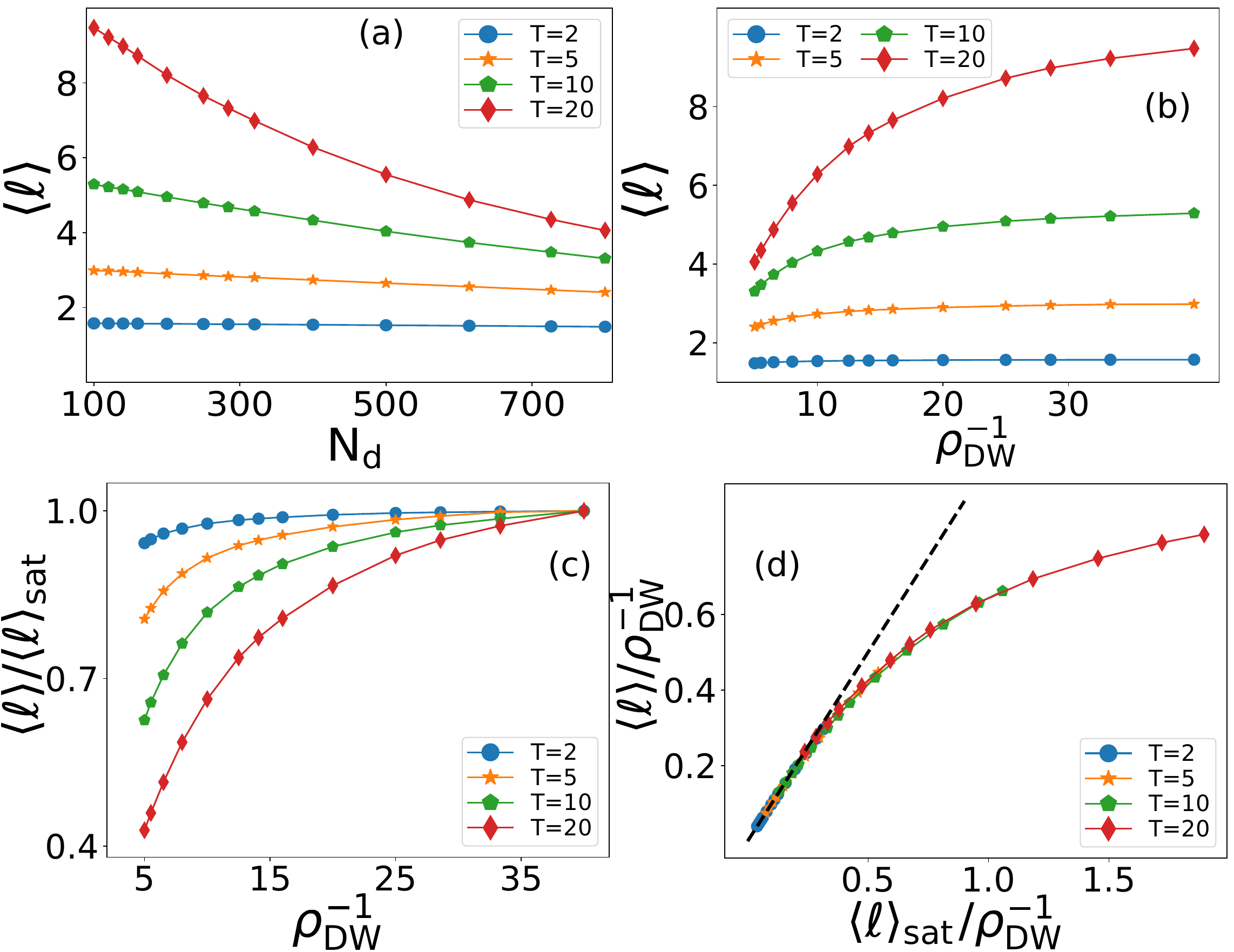}
\caption{(a) Average length, $\al$, of doublons in equilibrium as a function of $N_d$. (b) $\al$ as a function of $\rho_{DW}^{-1}$. $\al$ saturates to a $T$-dependent value, $\al_\text{sat}$, when $\rho_{DW}^{-1}$ is large. (c) $\al/\al_\text{sat}$ is shown as a function of $\rho_{DW}^{-1}$. (d) The plot of $\al/\rho_{DW}^{-1}$ as a function of $\al_\text{sat}/\rho_{DW}^{-1}$ follows a master curve for data at different $T$, confirming our scaling ansatz, Eq. (\ref{fssansatz}).
	The dashed line has slope = 1, as predicted by our scaling ansatz, Eq. (\ref{scalingansatz}).	
System size $N=4000$ for all the simulation data presented in this figure. Each point represents an average over $10^4$ realisations.}
\label{ssdl_scaling}
\end{figure}

\subsection{Effect of domain wall density on domain lengths}
There are two competing length scales in the system in equilibrium: one is determined by $T$ and the other by the domain wall density $\rho_{DW}$.
When the number of DWs, $N_d$, is large, $\rho_{DW}^{-1}=N/N_d$ gives the average domain size, and the equilibrium length $\al$ set by $T$ cannot be larger than $\rho_{DW}^{-1}$. On the other hand, in the limit $\al\ll\rho_{DW}^{-1}$, DW density is not relevant. To study this effect, we place $N_d$ number of DWs in a system of size $N$ and study $\al$ with varying $N_d$ at different $T$. Figure \ref{ssdl_scaling}(a) shows $\al$ at four values of $T$ as a function of $N_d$, $\al$ decreases with increasing $N_d$. Figure \ref{ssdl_scaling}(b) shows $\al$ as a function of $\rho_{DW}^{-1}$; as evident from the plot, $\al$ saturates to $\al_\text{sat}$, a $T$-dependent value when $\rho_{DW}^{-1}$ is large enough. We find $\al_\text{sat}$ increases with $T$. Plot of $\al/\al_\text{sat}$ as a function of $\rho_{DW}^{-1}$, as presented in Fig. \ref{ssdl_scaling}(c), shows the approach to saturation depends on $T$, which can be characterized in terms of $\al_\text{sat}$.

To summarize these observations, we propose the following scaling form for $\al$:
\begin{equation}\label{fssansatz}
\al=\rho_{DW}^{-1}\mathcal{F}(\al_\text{sat}/\rho_{DW}^{-1}),
\end{equation}
with the function $\mathcal{F}(x)$ having the following properties
\begin{equation}\label{scalingansatz}
\mathcal{F}=
\begin{cases}
x, \text{ when } x \to 0\\
1, \text{ when } x\to\infty.
\end{cases}
\end{equation}
Therefore, the plot of $\al/\rho_{DW}^{-1}$ as a function of $\al_\text{sat}/\rho_{DW}^{-1}$ for different sets of data should follow a master curve. The data collapse, shown in Fig. \ref{ssdl_scaling}(d), is thus consistent with our ansatz, Eq. (\ref{fssansatz}).

\section{Results: dynamical properties}
\label{results2}
%\subsection{Ballistic domain wall dynamics}
%\label{ballisticDW}

In this section, we demonstrate that the dynamics of the DWD model can
be understood in terms of the motion of domain walls (DWs) and
doublons. We first summarize the contents of the section.

In Sec. \ref{single_dw_dynamics}, we derive the drift velocity and diffusion constant of a single DW, while in Sec. \ref{single_doublon_dynamics} we study the dynamics of a pair of DWs. Distant DWs tend to approach each other and form a doublon.
Subsequently, the doublon performs a random walk and we derive
its diffusion constant.

In Sections \ref{auto_corr} (mainly numerical results) and \ref{auto_corr_analytic} (analytic estimates) we study the autocorrelation function in the DWD model with a
macroscopic number of DWs. We show that the early time behaviour is determined by DW motion, while the late time decay $(~t^{-1/2})$ is determined by the dynamics of a doublon fluid comprised of compressible particles with exclusion interactions. Interestingly, the crossover from a DW-dominated to a doublon-dominated regime occurs at a time $t_c$ which varies non-monotonically with temperature $T$. At low $T$, there is a large time scale for individual spin flips, leading to a late onset of the doublon power-law regime. At high $T$, the situation is more interesting, in that there is a large regime of time in which a stretched exponential decay precedes the power-law decay coming from doublon dynamics.

%\r{\subsection{Diffusion Constant of Domain Wall and Doublon}
%To look into the dynamics, we firstly looked into the dynamical behaviour of a single domain wall as shown in Fig. \ref{diffusion_cons}. A change in the slope of Fig. \ref{diffusion_cons} (a) indicates a change in the dynamics i.e. from dynamics of a single domain wall to that of a doublon. After doing some analytic calculations we found that the diffusion constant of the DW ($D_{DW}$) will go as $(1+c)$ whereas that of doublon ($D_{Doublon}$) will go as $c$ (See Fig. \ref{diffusion_cons} (b) and (c)).}

\begin{figure*}
	\includegraphics[width=16.6cm]{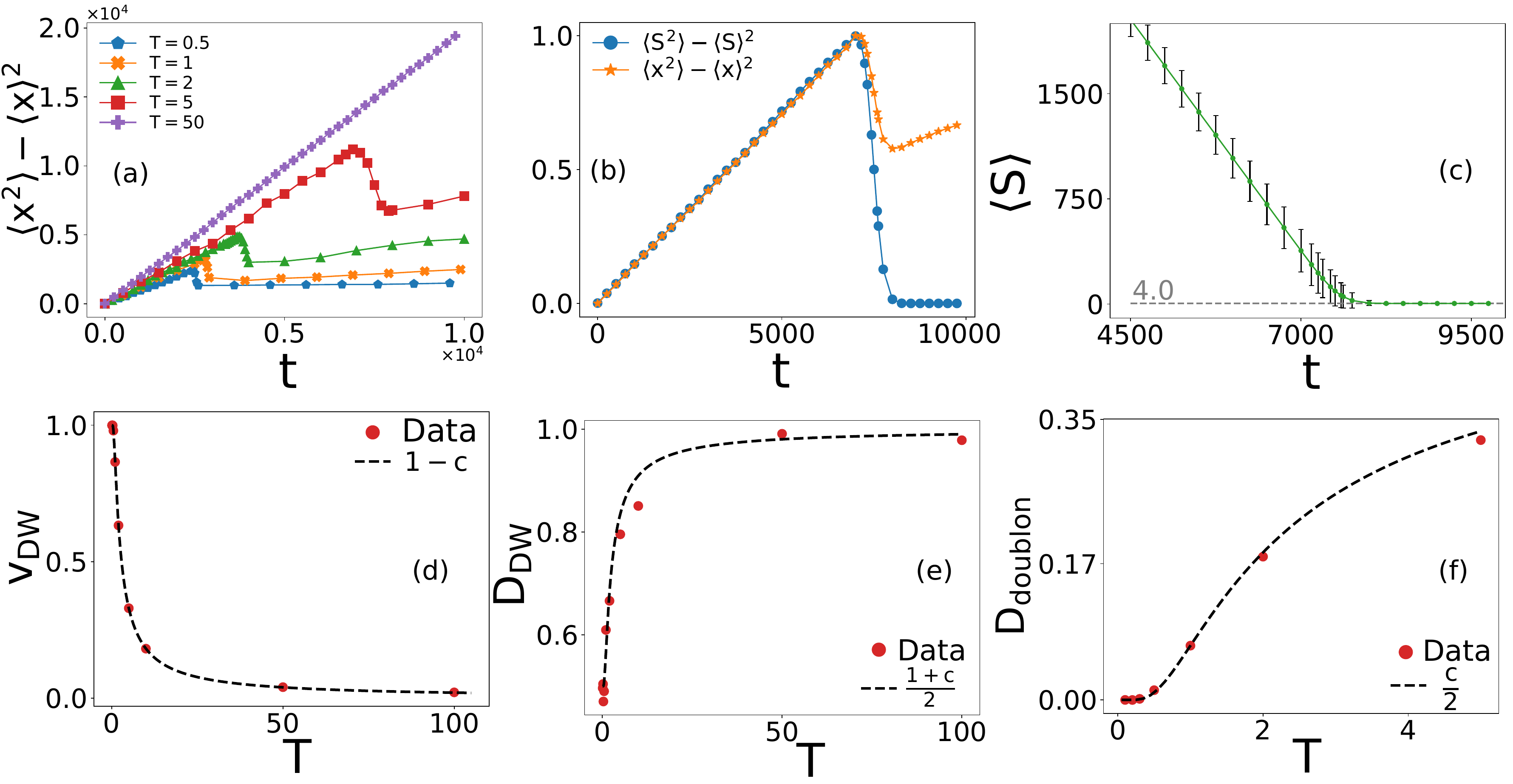}
	\caption{(a) Variance of the position of DW ($\langle x^2 \rangle - \langle x \rangle^2$) vs t for different temperatures. The behaviour shows a kink at some intermediate time which changes with temperature. This indicates a change from domain wall dynamics to doublon dynamics. As $T$ increases the crossover time also increases. The slope of the initial part of $\langle x^2 \rangle - \langle x \rangle^2$ gives $D_{DW}$, whereas that after the kink gives $D_{doublon}$. (b) We show the variance of the separation between the two DWs ($\langle S^2 \rangle - \langle S \rangle^2$) and $\langle x^2 \rangle - \langle x \rangle^2$. We scaled their values such that their maxima become one. At long times, $\langle S^2 \rangle - \langle S \rangle^2$ saturates, whereas $\langle x^2 \rangle - \langle x \rangle^2$ continues to grow. (c) $\langle S \rangle$ decreases with $t$ and then saturates to a non-zero value when the doublons form. The error bars is the standard deviation of $S$, which shows a non-monotonic behaviour [$T = 5$ for (b) and (c)]. (d) Domain wall velocity, $v_\text{DW}$, as a function of $T$ follow $v_\text{DW}=1-c$, where $c=\exp(-2/T)$. Symbols are simulation data, and the line is theory. (e) $D_\text{DW}$ is given by $(1 + c)/2$ (Eq. \ref{dw_diff}). 
(f) $D_\text{doublon}$  is given by $c/2$. [$N = 10^4$, $N_d = 2$ and initial separation is $N/2$].}
	\label{diffusion_cons}
\end{figure*}

\begin{figure*}
	\includegraphics[width=16.6cm]{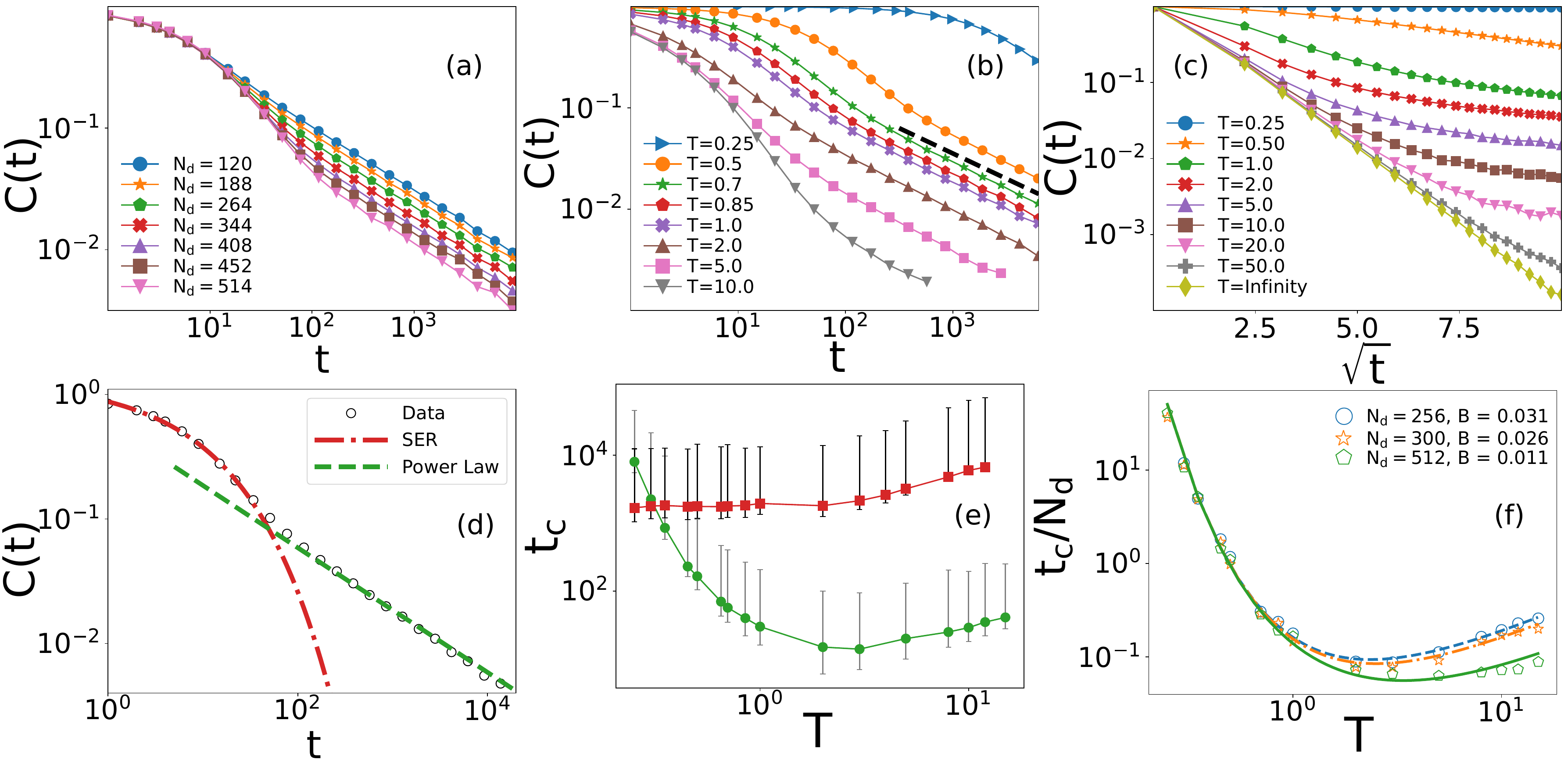}
	\caption{(a) Decay of the autocorrelation function, $C(t)$ [Eq. (\ref{coftdef})], for different number of DWs, $N_d$, at $T=1$. (b) $C(t)$ as a function of $t$ at different $T$ and $N_d=344$. The dashed black line shows a power law with exponent $1/2$. (c) $C(t)$ vs $\sqrt{t}$ plot implying the behaviour that the stretched exponential decay corresponds to that of $T = \infty$ ($N_d = 344$) (d) Decay of $C(t)$ shows two different regimes: it follows an SER at short $t$, the fitted function is $f(t)\sim \exp[-(t/\tau)^{1/2}]$, and a power-law at long times, the fitted function is $f(t)\sim t^{-1/2}$. We have used $N_d=344$ and $T=1.0$. (e) The crossover-time, $t_c$, shows a non-monotonic trend (circles) as most MC attempts are unsuccessful at low $T$. The trend becomes monotonic (squares) when we define time via the accepted moves alone. The error bars represent an estimate of the uncertainty in evaluating $t_c$ and $N_d=344$. (f) $t_c/N_d$ collapses to a single curve at low $T$ and then deviates at high $T$ for different values of $N_d$ [$N=1024$ for Figs. (a)-(f)].}
	\label{autocorrfn}
\end{figure*}

\subsection{Single domain wall: dynamics}
	\label{single_dw_dynamics}
	 The dynamical behaviour of a single domain wall is isomorphic to that of a random walker which in each time step moves right with probability $p$, left with probability $q$ and stays put with probability $w = 1 - p - q$. The probability of $l$ left steps, $r$ right steps, and $s$ stationary steps is
	\begin{equation}
		P(l,r,s) = \frac{N!}{l!r!s!}p^rq^lw^s.
	\end{equation}
Let us denote the displacement after $N_t$ steps by $n=(l-r)$. It is straightforward to see that $\langle n \rangle = N_t(p-q)$ while the variance is $\sigma_n^2 \equiv \langle n^2 \rangle - \langle n\rangle^2 = N_t[p(1 - p) + q(1 - q)]$. Considering up spins in the right of the DW, the rates of the DW to step right or left are $1$ and $c$, let us associate $p$ and $q$ with probabilities for the DW to move right or left in a small time $\delta t = 1/N$ where $N \equiv $ system size. Evidently, after $N_t = N$ steps (1 Monte Carlo step), we have $\langle n \rangle = (1 - c)$ and $\sigma_n^2 = (1 + c)$ where we have dropped terms of order $(\delta t)^2$. This implies that the mean velocity of the DW is 
	\begin{equation}
		\label{speed}
		v_{DW} = 1 - c,
	\end{equation} and the diffusion constant is 
	 \begin{equation}\label{dw_diff}
	 	D_\text{DW} =\frac{\langle \sigma^2\rangle}{2}=\frac{1 + c}{2}.
	 \end{equation}
 
  A confirmation of Eq. (\ref{speed}) is obtained by studying the ballistic motion of a DW in a system of size $l$ + 1, where the left most spin is $\downarrow$ and the rest are $\uparrow$ (without periodic boundary condition). Simulations of these systems show that the time taken to flip the last but one spin increases linearly with $l$ with slope $1/v_{DW}$. Figure \ref{diffusion_cons}(d) shows that $v_\text{DW}$, determined from simulation, agrees well with Eq. \ref{speed}.

\subsection{Single doublon: formation and diffusivity}
	\label{single_doublon_dynamics}
	Consider an initial state with two large domains, of $\uparrow$ and $\downarrow$ spins of size $N/2$ each. Periodic boundary conditions imply that there are two well-separated DWs. They attract each other and ultimately form a doublon which itself diffuses in the system after its formation. To study the formation and dynamics of a single doublon, we started with a system with $N = 10^4$. Since $D_{DW}$ and $D_{doublon}$ differ substantially, there is a sharp fall of the variance of the position of a DW (Fig. \ref{diffusion_cons} (a)), indicating the formation of a doublon which then move diffusively. To confirm this we looked into the variance of the separation between two DWs ($\langle S^2 \rangle - \langle S \rangle^2$)  Fig. \ref{diffusion_cons} (b), which saturates to a non zero value after the kink is obsereved in Fig. \ref{diffusion_cons} (a), indicating the formation of a doublon. Fig. \ref{diffusion_cons} (c) shows that even though $S$ has a monotonic behaviour, its standard deviation is non-monotonic. The process is shown pictorially in Fig. (\ref{schematic_dyn}).

	We now obtain the diffusivity of a doublon, following a similar argument as for a single DW, except the values of  $p$, $q$, and $w$ are different. We consider the low-$T$ limit $(c \ll 1)$ in which case the doublon is basically a single $\uparrow$ spin bracketed by two DWs. The doublon can move only when one of the two DWs move outwards by flipping a $\downarrow$ spin, e.g. it can go from $\downarrow\downarrow\boldsymbol{\uparrow}\downarrow\downarrow$ to $\downarrow\downarrow\boldsymbol{\uparrow}\boldsymbol{\uparrow}\downarrow$ with probability $c\delta t$. Next, on a much faster time scale, one of the two $\uparrow$ spins will flip, resulting in either the initial state $\downarrow\downarrow\boldsymbol{\uparrow}\downarrow\downarrow$ or state $\downarrow\downarrow\downarrow\boldsymbol{\uparrow}\downarrow$, with equal probability. Thus the probability of moving a doublon one step right in a small time $\delta t$ is $p=c\delta t/2$. Likewise the probability of moving it one step left is $q=c\delta t/2$. Thus the variance of the doublon motion after $N$ steps is $c$, implying the diffusion constant 
	\begin{equation}\label{doublon_diff}
		D_\text{doublon} = \frac{c}{2}
	\end{equation}

 We now test Eqs. (\ref{dw_diff}) and (\ref{doublon_diff}) in our simulations. Figure \ref{diffusion_cons}(a) shows the plots of $(\langle x^2\rangle-\langle x\rangle^2)$ as a function of time. The initial slope gives the diffusivity of the DWs and the latter slope gives that of the doublons. Figures \ref{diffusion_cons}(e) and (f) show $D_\text{DW}$ and  $D_\text{doublon}$, respectively; the symbols are simulation results and the lines are the theoretical predictions. Although Eq. (\ref{doublon_diff}) was derived in the limit of low $T$, the numerical results suggest a wider range of validility.

\subsection{Autocorrelation function in steady state}
\label{auto_corr}
We now study the DWD model with a macroscopic number of DWs and show that relaxation at short times can be understood via the dynamics of DWs, while the long-time relaxation involves the diffusion of doublons. 

The relaxation dynamics is characterized by the autocorrelation function, $C(t)$, defined as
\begin{equation}\label{coftdef}
C(t)={ \f{1}{N}\sum_{i=1}^N \langle S_i(t_0)S_i(t+t_0)\rangle}- S_{avg}^2,
\end{equation}
where $S_{avg}$ is the average equilibrium magnetization and $\langle \ldots\rangle$ define averages over initial configurations.

%$\langle \ldots\rangle_{t_0}$ and the overline define averages over the initial time $t_0$ and initial configurations, respectively. 

We show the behaviour of $C(t)$ at $T=1$ with different numbers of DWs in Fig. \ref{autocorrfn}(a) and with a fixed number of DWs, $N_d=344$, in a system with $N=1024$ at various $T$ in Fig. \ref{autocorrfn}(b). Two distinct types of decay are apparent from the figures: the short-time decay is relatively rapid while the long-time behaviour follows a power law, $t^{-1/2}$. At relatively high $T(>1)$ the short time decay follows a stretched exponential, $\exp[-(t/\tau)^{\beta}]$ with $\beta=1/2$ (Fig. \ref{autocorrfn}(c)), while the long time decay is governed by doublon diffusion and follows a power law, $t^{-1/2}$. Fits to these specific functions are shown in Fig. \ref{autocorrfn}(d). Reference \cite{Causer2020} did not study the long-time behaviour of the auto-correlation function and hence did not report this crossover to the power-law behaviour.

As discussed earlier, the $T=\infty$ limit of the system is equivalent to the diffusive dynamics of DWs \cite{skinner1983,spohn1989}, and $C(t)$ decays via SER. At finite $T$, there is a tendency of the DWs to move inwards into an $\uparrow$ spin cluster. However, the effect is negligible at short times, and the DW dynamics remains diffusive, implying SER of $C(t)$ (see \ref{auto_corr_analytic} below).
 %Note that many possible moves are unsuccessful at low $T$: this affects the time scale of the dynamics. Interestingly, when we define units of time as $M$ number of accepted moves, we find a monotonic behaviour for $t_c$, as shown in Fig. \ref{autocorrfn}(f). Thus, the non-monotonic behaviour has its origin in the rejected moves of MC dynamics at low $T$.

{\it Estimation of crossover time:}
	The crossover time, $t_c$, between short time decay and the long-time power-law decay is seen to be non-monotonic in $T$. At low $T$, many possible moves are unsuccessful so that $t_c$ increases with decreasing $T$; on other hand, at high $T$, $t_c$ increases with increasing $T$, since the bias decreases with increasing $T$.
	
	Interestingly, if we define units of time as the number of accepted moves, we find a monotonic behaviour for $t_c$, as shown in Fig. \ref{autocorrfn}(e). Thus, the non-monotonic behaviour has its origin in the rejected moves of MC dynamics at low $T$.
	
	Since the crossover results from the change in the nature of the excitations from DW to doublon, we estimate the crossover time as follows. At low $T$, doublons are already tightly bound, and dynamics proceeds mostly via the flip of down spins. The rate of these spin-flips is $\exp(-2/T)$, and most attempts are rejected at low $T$. Therefore, $t_c$ should increase as $\exp(2/T)$ with decreasing $T$. On the other hand, at high $T$, the average domain length saturates to a value that depends on $N$ and $N_d$ (Fig. \ref{mfpt_domain}b). In this regime, we estimate $t_c\sim \al/v_\text{DW}= \al/(1-c)$. We use an interpolation form $t_c\sim a/c+b/(1-c)$ where $a$ and $b$ should depend on $N_d$ and $N$. At low $T$, doublons are typically of one spin width. If we chose one of these up spins, it cannot flip due to the dynamical rule of the DWD model. For a system of $N$ spins, such choice is proportional to $N_d$. Thus, we expect $a=A N_d$, where $A$ is a constant. Therefore, we plot $t_c/N_d$ in Fig. \ref{autocorrfn}(f) for three different values of $N_d$. A fit of the form $t_c/N_d=A/c+B/(1-c)$ gives $A=0.017$ and different values of $B$ (Fig. \ref{autocorrfn}f).

\subsection{Analytic arguments for the crossover of autocorrelation function}
\label{auto_corr_analytic}
We now present approximate analytic arguments relating the excitation dynamics to the decay of the autocorrelation function in the DWD model.

The motion of the DWs has both diffusive and ballistic components. The latter results from a bias in the movement of the DWs towards the $\uparrow$ spins. Individual DWs move diffusively at short times, while the bias becomes significant at longer times, leading to the formation of a bound pair of DWs i.e., doublons, which then move together diffusively. Thus, at short times, $t\ll t_c$, the diffusive motion of DWs governs the decay of $C(t)$. In this case, the relaxation scenario proposed by Spohn is valid \cite{spohn1989}: Since the dynamics is constrained such that only DWs can move, a segment of up (or down) spins can relax only via the movements of the DWs. The probability of having a large segment of length $L$ of up spins is exponentially small and proportional to $\exp[-f L]$, where $f$ is a free energy scale. Since the dynamics of the DWs is diffusive, the time $t$ a DW takes to move this distance is of the order of $L^2$. Thus, the spin-spin autocorrelation function decays as $\exp[-\sqrt{t/\tau}]$. Upper and lower bounds for $\tau$ were derived in Ref. \cite{spohn1989}.

\begin{figure}
	\includegraphics[width=7.6cm]{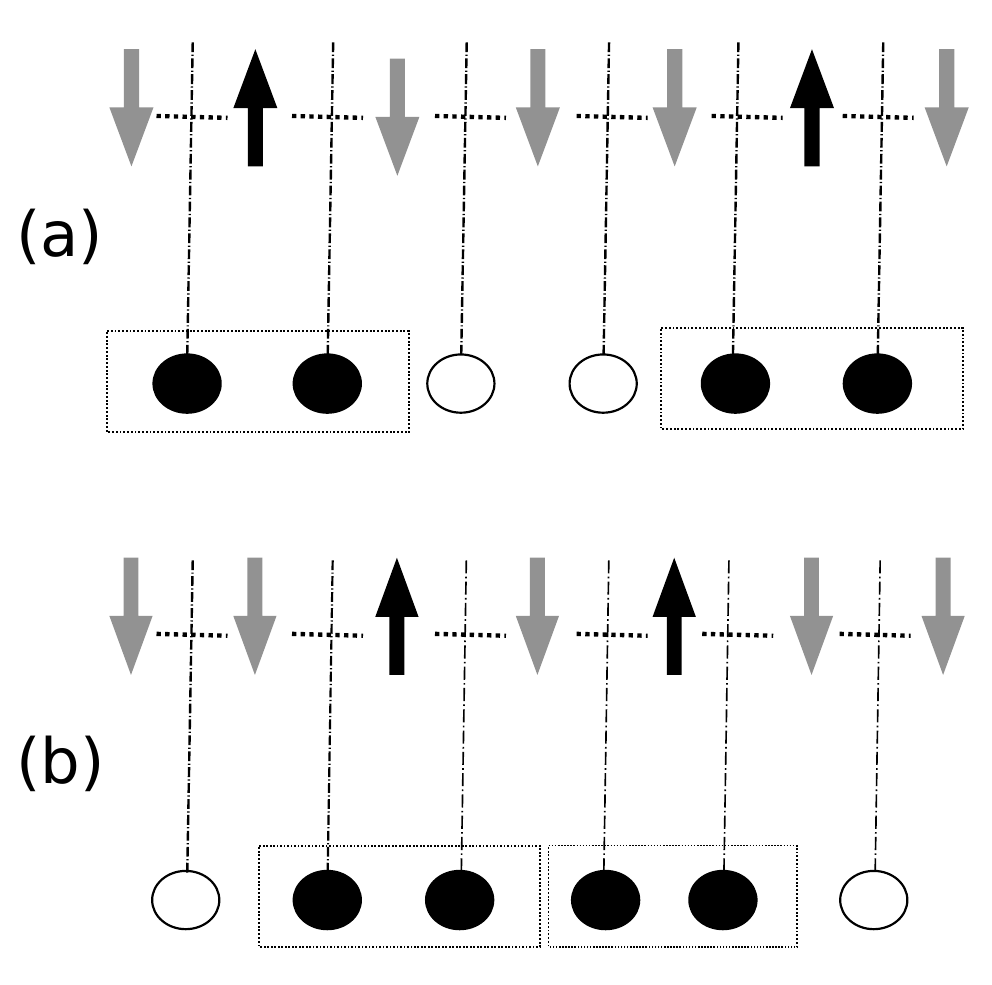}
	\caption{Mapping of the DWD model at low $T$ to a system of dimers. (a) The DW between two opposite spins in the DWD model is represented by a particle, while the absence of a DW between two like spins is represented by a hole. In this representation, a dimer represents a doublon. (b) The fact that two DWs cannot be closer than two lattice spacings implies a hard core interaction between dimers.}
	\label{dimers}
\end{figure}

On the other hand, on a time-scale $t\gtrsim t_c$, the bias in the DW movement is significant, and a pair of DWs can bind as a doublon; further relaxation of the system can be described in terms of the dynamics of doublons. We argue below that the diffusive dynamics of the doublons leads to a power-law relaxation in the low $T$ limit. In this limit the doublon consists of tightly bound DWs around a single $\uparrow$ spin. Hard core interactions between DWs imply that the distance of closest approach between the centres of two doublons is two lattice spacings. It is straightforward to see that there is one to one mapping between the doublon problem and the dynamics of diffusing hard dimers in 1D (Fig. \ref{dimers} (a)). In turn the latter problem shares the elements of exclusion and diffusion (Fig. \ref{dimers} (b)) with the simple exclusion process \cite{Menon1997,stinchcombe1993,liggettbook,shamik2011}, so we expect the long time behaviours to be similar. Consequently, $C(t)$ is expected to decay as $t^{-1/2}$ at long enough times; this agrees well with the numerical results presented in Fig. \ref{autocorrfn}.

%The number of doublons in our system is constant and equals $N_d/2$.
 %\b{ We now argue that in the limit of large system size, our system resembles the simple exclusion process \cite{stinchcombe1993,spitzer1970,liggettbook,shamik2011}. First, the doublons cannot cross each other, as DWs do not cross themselves. Second, the closest distance that the doublons can come is two lattice-spacings apart and not next to each other. But, at large length and time scales, the distinction from simple exclusion should not matter. Third, at low $T$, the doublons are very tightly bound objects: they have typical size of one spin. }
 %	\r{In this limit, we can map the DWD system to a system of dimers (Fig. \ref{dimers}). 
 %	 We place a particle in between two opposite spins and hole between two like spins. A doublon is then represented by a dimer. We show a typical configuration in Fig. \ref{dimers}(a). The minimum distance between two consecutive doublons in the DWD model can be one spin apart. However, in the dimer representation, they can come next to each other, as shown in Fig. \ref{dimers}(b). Reference \cite{Menon1997} have shown that the dynamics of a system of hard-core dimers can be mapped to that of a simple exclusion process. 
 %	 Therefore, we can study the doublon dynamics by treating them as hard-core particles in the system, resembling a simple exclusion process. Consequently, $C(t)$ is expected to decay as $t^{-1/2}$ at long enough times; this agrees well with the numerical results presented in Fig. \ref{autocorrfn}. }

\section{Discussion and Conclusion}
\label{disc}
Crossover scenarios are ubiquitous in many complex systems \cite{metzler2000,metzler2002,vaibhav2020,ganazzoli2002}, including biological systems, where relaxation dynamics can have additional nontrivial contributions from activity \cite{clara2020,souvik2020}. The change of the autocorrelation function generally signifies a change of the mechanism operating at various times. Below, we illustrate this by summarising our results for the one-dimensional domain wall to doublon (DWD) model, and also discussing SER to power-law crossovers in a number of other systems.

\begin{itemize}

\item 
The DWD model falls in the broad category of kinetically constrained models, as a spin can only flip when its nearest neighbors are of opposite signs. A uniform magnetic field induces a rich behaviour, due to attraction between pairs of DWs leading to the formation of doublons. At short times, the system relaxes due to the motion of individual DWs, while at long times it relaxes via the diffusive motion of doublons, comprised of a number of up spins between two DWs. In the steady state, the distribution of doublon sizes was shown to fall exponentially, with a $T$-dependent length scale. 
The simplicity of the model allows the decay of the autocorrelation function to be understood. The $T=\infty$ limit of the model is tantamount to the energy-conserving dynamics of an Ising model at zero temperature, which is known to show SER \cite{spohn1989}. The same form describes the relaxation at short times in the DWD model at high $T$. At later times, pairs of DWs form doublons, which diffuse with a hard core constraint, leading ultimately to a power-law decay of the autocorrelation function.

\item 
A conserved number of DWs also arises in a model with competing nearest neighbour and next nearest neighbour interactions, when quenched to $T=0$ \cite{vaibhav2020}. The spin-spin autocorrelation function shows SER with an unusual feature, namely a relaxation time which depends on system size. On averaging over random initial conditions, the autocorrelation function has the Mittag-Leffler form \cite{mainardi2020} which interpolates between a stretched exponential and power law, characterized by the same exponents.

\item 
The Mittag-Leffler function arises in several problems involving fractional dynamics. An example is Ref. \cite{metzler2002}, where it arises in trap models having a power-law waiting time distribution.

\item 
Such a crossover also arises in the survival probability of a random walk on a fractal lattice with spatially correlated traps \cite{plyukhin2016}. The exponents in this case depend on fractal dimensions of both the host lattice and the sublattice comprised of the traps.

\item
Systems which show glassy behaviour also manifest a similar crossover. In recent work \cite{pareek2023}, the long-time power-law behaviour of $Q(t)$ was shown to follow $\sim t^{-d/2}$, where $d$ is the dimension, which is similar to the DWD model. However, the SER behaviour is different: whereas the stretching exponent in the DWD model is $T$-independent, that for $Q(t)$ depends on $T$. Further, in a system of non-entangled polymers \cite{ganazzoli2002}, the intermediate scattering function was found to deviate from a stretched exponential, which as noted in \cite{pareek2023}, implies that $Q(t)$ shows a crossover from SER to power-law. Similar crossover for $Q(t)$ arises in simulations of a confluent monolayer (Fig. 5(e) in \cite{souvik2020}), and separately also in simulations of active probes in collidal glasses (Fig. S12(b) in \cite{clara2020}).
\end{itemize}

As we see from the discussion above, the SER-power-law crossover occurs in a wide variety of systems. The advantage of the DWD model is that it allows for a detailed understanding in terms of a change of the predominant excitations from DWs to doublons. 

\section{Acknowledgements}
We thank Juan P. Garrahan and Eli Barkai for useful correspondence. M.B. acknowledges support under the DAE Homi Bhabha Chair Professorship of the Department of Atomic Energy, India. We acknowledge the support of the Department of Atomic Energy, Government of India, under Project Identification No. RTI 4007. S.K.N. thanks SERB for grant via SRG/2021/002014.

\bibliography{crossover.bib}
\bibliographystyle{apsrev4-1}

\end{document}